**Large-Scale and Robust Multifunctional Vertically-Aligned MoS$_2$ Photo-Memristors**

Kamalakannan Ranganathan, Mor Feingenbaum and Ariel Ismach[#]

Department of Materials Science and Engineering, Tel Aviv University, Ramat Aviv, Tel Aviv, 6997801, Israel
#E-mail: aismach@tauex.tau.ac.il

**Abstract**

Memristive devices have drawn considerable research attention due to their potential applications in non-volatile memory and neuromorphic computing. The combination of resistive switching devices with light-responsive materials is considered a novel way to integrate optical information with electrical circuitry. On the other hand, 2D materials have attracted substantial consideration thank to their unique crystal structure, as reflected in their chemical and physical properties. Although not the major focus, van der Waals solids were proven to be potential candidates in memristive devices. In this scheme, the majority of the resistive switching devices were implemented on planar flakes, obtained by mechanical exfoliation. Here we utilize a facile and robust methodology to grow large-scale vertically aligned MoS$_2$ (VA-MoS$_2$) films on standard silicon substrates. Memristive devices with the structure silver/VA-MoS$_2$/Si are shown to have low set-ON voltages (<0.5V), large-retention times (>2x10$^4$ s) and high thermal stability (up to 350 °C). The proposed memristive device also exhibits long term potentiation / depression (LTP/LTD) and photo-active memory states. The large-scale fabrication, together with the low operating voltages, high thermal stability, light-responsive behaviour and long-term potentiation/depression, makes this approach very appealing for real-life non-volatile memory applications.

Keywords: Vertically Aligned MoS$_2$, memristor, electrochemical metallization, photomemristor, memory states, potentiation

**Introduction**

The search of materials for robust memory storage has been a prime focus for developing future computing capabilities.[1] Among non-volatile memory storage techniques, the memristor is one of the leading potential candidates for replacing or complementing CMOS technologies.[2] Memristors were found to exhibit multiple analogue states for information processing as well.[3]

These makes the memristors a potential candidate for solving the long standing problem known as 'memory wall' of the von-Neumann architecture by using data storage and data processing from the same device.[3a, 4] The standard memristor configuration complies a metal oxide thin film sandwiched between two metal contacts (MIM).[2b, 5] The resistive switching mechanism in these type of devices depends on the metal contact type, and they are classified in two main groups, active and non-active metal contacts. When the later type of contacts is used (such as gold), the low-resistance mode (LRS) is achieved by the formation of a conductive path by the migration and accumulation of oxygen anions.[2a, 6] When an "active" metal is used as one of the contacts, such as copper and silver, the high conductance state is reached by the diffusion of the Cu or Ag ions to form a conductive filament,[6b] or to reduce locally the resistance across the solid electrolyte, without necessarily forming a continuous metallic wire.[1b, 2b] Both processes are highly stochastic.

Atomic-thin semiconductors, such as the transition metal dichalchogenide (TMDC) family have attracted great scientific and technological attention due to their interesting physical and chemical properties, which make them promising candidates in a wide range of applications, such as nanoeletcronics,[7] optoelectronics,[8] catalysis[9] and energy storage.[9c-e, 10] Such atomic crystals were also suggested as solid electrolytes in resistive switching devices.[1a, 11] To our knowledge, only *planar* single- and few-layer 2D materials were implemented into memristor devices with top and bottom metal contacts. Therefore, the conductive path in these cases, is formed perpendicular to the basal plane, along defects present in the layers, such as sulfur vacancies. Despite these promising device prototyping, mostly based on mechanically exfoliated flakes, the large-scale single- and few-layer TMDCs formation with the desired homogeneity, as required for the fabrication of such memory devices, is still a great challenge.[12] In addition, making a vertical device with the structure metal - 2D material - metal, is prone to malfunction due to electrical leaks caused by intrinsic defects and the atomic size of such layers. Vertically aligned (VA) TMDCs were shown to be formed by the calchogenization of transition metal thin films or foils.[9a, 13] Thin films of VA-$MoS_2$,[9a, 10, 13a, 13b] -$MoSe_2$[13b] and -$WSe_2$[13c] were reported and their electronic and catalytic properties tested. These films therefore, offer some interesting features, not available on planar few-layer TMDCs. First, the sulfurization/selenization process is scalable and large-area thin films of VA-TMDCs can be realized in a consistent manner. Second, the VA-TMDC configuration can be advantageous in a vertical memristive device due to the enhanced ionic/cationic diffusion in between vdW layers.[9a, 9d, 9e, 10] Furthermore, such VA-TMDCs can in principle be doped/alloyed *in-situ* during the sulfurization process, therefore, adding functionality to the films.[9a, 14]

Conventional memristor devices based on a metal oxide thin film as the active layer, are not expected to be light sensitive due to their wide band-gap and disordered nature (amorphous) of such layers. Therefore, the integration of metal-oxide memristors into light sensitive devices is not feasible. The integration of optical sensing with electrical circuitry is very interesting for future energy-efficient electronic systems.[15] TMDCs-based photodetectors have shown fast response and high responsivities,[8, 15] therefore, allowing to integrate optical sensing and data storage. Indeed, photo-sensitive memory devices were demonstrated with TMDC-based planar configuration.[15-16] However, as stated above, the vast majority of such reports were performed on exfoliated flakes as proof of concept studies.[15-16] Moving forward to real life applications requires a highly consistent and robust synthetic methodology.

An additional interesting functionality is the ability of electrical components to operate at high temperatures, often defined as such higher than 150 °C, as is considered to be the temperature in which electrical performance is highly degraded.[17] To our knowledge, research on high temperature memristor performance is very scarce. For example, the performance at temperatures up to ~180 °C and ~860 °C were tested in resistive switching devices based on SiCN[18] and $HfO_x$[19] thin films, respectively. Graphene/$MoS_{2-x}O_x$/graphene layered devices were shown to operate up to ~340 °C.[11d] Hence, the development of memristive devices at high operation temperatures are desirable to expand their functionality and be used in harsh conditions.

In this study we demonstrate the successful fabrication of large-scale VA-$MoS_2$ based memristors with an active metal contact (top contact-Ag) and a heavily doped Si wafer as the bottom electrode. Such resistive switching devices exhibit low SET voltages (<0.5 V) and good stability at high temperatures (up to 350 °C), to our knowledge, the higher measured so far. The Ag/VA-$MoS_2$/Si devices are shown to be light sensitive, thus allowing for future integration between light-stimulation and data storage. The memristive device also exhibits long term potentiation / depression, as needed for neuromorphic computing applications. The work presented here may allow for the large-scale fabrication of functional (photo) resistive switching devices based on VA-TMDCs, with high stability in high temperatures.

**Result and discussion**

The VA-MoS$_2$ thin films were prepared by the sulfurization of a pre-deposited 15 nm-thick Mo film on Si substrates (see Figure S1 and experimental section for details). Here, heavily doped Si wafer was chosen to serve as the growth substrate and the bottom contact without any additional metal fabrication. Prior to the Mo film deposition, the native oxide Si wafer is treated with an HF solution in order to assure no SiO$_2$ is present at the metal evaporation step. Figure 1 shows the film characterization. The in-plane TEM image, Figure 1(a), was taken after transferring the MoS$_2$ film to a standard holey carbon TEM grid. The MoS$_2$ standing layered structure is clearly seen. The measured interlayer distance of ~0.6 nm is consistent with the MoS$_2$ bulk crystal structure.[13a] To visualize the interface between the MoS$_2$ and the Si substrate, as well to further corroborate the vertical alignment of the layers, cross section samples were prepared using FIB-related methods, as specified in the experimental section. Figure 1 (b) shows the TEM image of such samples, in which the VA-MoS$_2$ is noticed.[13a, 13b] The final thickness of the film is ~35nm, exhibiting a volume expansion of around ~2.2 from the initial 15 nm Mo film thickness, which is consistent with previous reports.[13a] A thin layer (3-5 nm thick) of SiO$_2$ is also present at the interface between the MoS$_2$ and the Si. The oxygen initially present at the surface of the Mo film and as residual species in the reactor chamber, must diffuse through the metal film and react with Si, due to its higher oxygen affinity, see the Ellingham diagram comparing between the free energy of formation for the SiO$_2$ and Mo-O phases, Figure S2. Nevertheless, this oxide layer is expected to have low density and porous structure, since the sulfurization temperature and gas composition does not fit the conditions required for a high quality dielectric layer formation.[20] Hence, the thin and porous SiO$_2$ interfacial layer is not expected to have a big influence on the memristor behaviour. HRTEM analysis in various spots indeed indicates a porous and amorphous with a thickness of less than 5 nm (Figures 1 (b) and S3). Previous reports on similar VA-MoS$_2$ films confirmed the presence of a layer at the interface as well.[13a, 20b] Raman spectroscopy, Figure 1 (c), shows the characteristic MoS$_2$ modes, the in-plane ($E^1_{2g}$) and out of plane ($A^1_g$) phonon modes were assigned at 383 cm$^{-1}$ and ~408 cm$^{-1}$, respectively. The peak at 520 cm$^{-1}$ arises from the Si substrate. The peak position difference $A^1_g - E^1_{2g}$, was found to be around 25 cm$^{-1}$, indicating the presence of a multi-layer MoS$_2$. In addition, the ratio $E^1_{2g}/A^1_g = 0.39$, reveals that the out-of-plane vibration mode, $A^1_g$, is more pronounced, reflecting the dominantly exposed MoS$_2$ edge sites, which are a signature of vertical growth.[9d, 21]

The chemical analysis was carried out by x-rays photoemission spectroscopy (XPS). The survey spectrum (Figure S4(a)) of the as grown MoS$_2$ shows the coexistence of Mo, S and O species on the surface of the film. High resolution XPS spectrum (Figure 1 (d)) of Mo 3d was resolved into three peaks. The peaks at 236.2 eV for $Mo^{6+}$ $3d_{3/2}$, 232.5 eV for $Mo^{4+}$ $3d_{3/2}$ and 226.6 for $Mo^{4+}$ $3d_{5/2}$ were assigned.[9b, 22] The survey spectrum of an $Ar^+$ Ion gun sputter cleaned surface (Figure S4(b)) indicates a drastic reduction of oxygen on the surface. Moreover, the $Mo^{6+}$ $3d_{3/2}$ peak also vanished (Figure S4(c)) after surface cleaning, indicating that the oxide species such as $MoO_3$ are mainly at the surface and are a result of processing the samples in atmospheric conditions. The $S^{2-}$ 2p (figure 1(e)) has been resolved into two peaks at 163.5 and 162.2 eV for $2p_{1/2}$ and $2p_{3/2}$, respectively. Further, XPS depth profile has been carried out to study the chemical and phase composition throughout the film. Figure 1(f) shows the result of such profiling. The system consists of a multi-layer structure of $MoS_2$ on top of a thin coated Si substrate. The high resolution XPS at the interface between Si and $MoS_2$ is referred to a thin layer of $SiO_2$ without any other phases, such as molybdenum silicide (Figure S5).

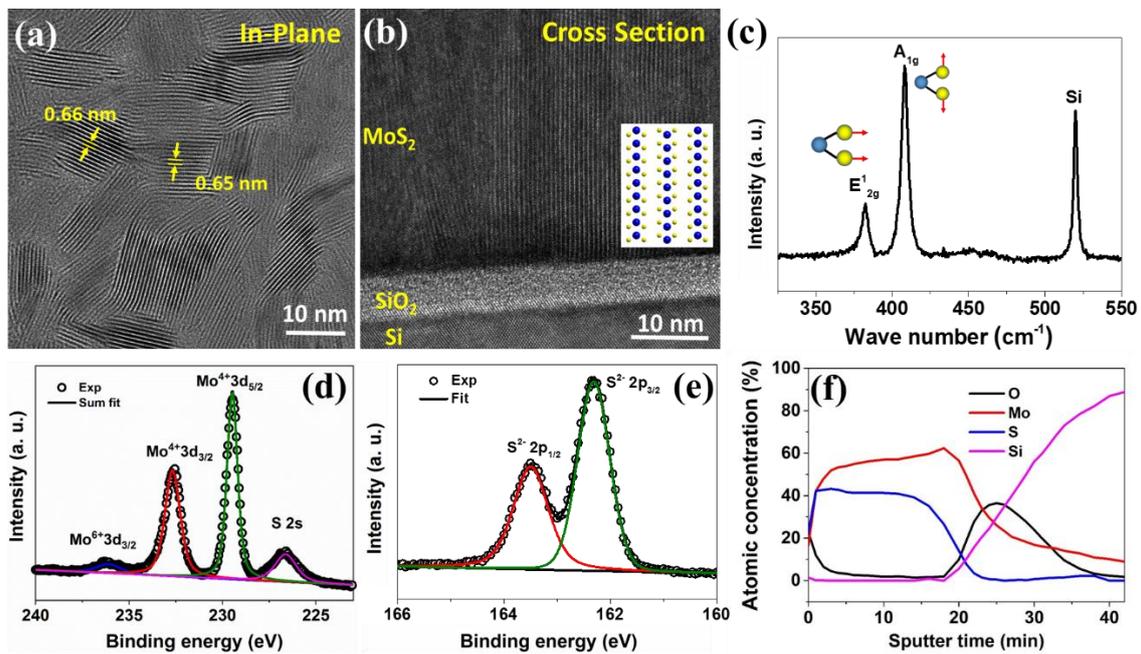

Figure 1: $VA-MoS_2$ films characterization: (a) In-plane HRTEM image showing the $MoS_2$ standing layers. (b) Cross-section HRTEM showing the vertically aligned $MoS_2$ layers, as well as the Si substrate and the interfacial $SiO_2$ thin layer. (c) Raman spectra showing the characteristic $MoS_2$ and Si modes. (d) - (e) high resolution XPS spectra for Mo 3d and S 2p, respectively. (f) XPS depth profile, atomic concentration as a function of sputtering time, revealing as well the presence of a $MoS_2/SiO_2/Si$ layered structure.

Figure 2 shows the electrical characterization of the VA-MoS$_2$ based memristor. A schematic representation of the measurement set-up is depicted in Figure 2 (a). Figures 2 (b)-(c) show the results obtained on a device made with a vertical structure of Au/MoS$_2$/Si. The current-voltage curve in Figures 2(b)-(c) exhibit the typical switching characteristics of such device, in which some hysteresis can be observed. On the other hand, a thin layer (30 nm) of Ag deposited in between the Au top contact and the VA-MoS$_2$ film, changed dramatically the device behaviour. First, a clear electroforming step, at 2.2 V, can be observed, Figure 2 (d)-(e), which is a common feature in a memristive device, where the conductive path across the solid electrolyte, the VA-MoS$_2$, is formed for the first time.[6b] The device shows a switching behaviour from high-resistance state (HRS, *i.e.*, OFF state) to the low resistance state (LRS, *i.e.*, ON state). Following the electroforming cycle, the device requires lower voltages to sweep (-0.5 V to 0.5 V) for the repeated switching form HRS to LRS and vice versa. While sweeping voltage from 0 to 0.5 the current increased drastically around ~ 0.45V, thus considered as the set voltage. Similarly, in the negative bias regime, the device current started a significant reduction around -0.3V, hence showing the reseting capability of the memory state. The reliability and robustness of the device was tested for more than 100 switching cycles, Figure S5, in which a low standard deviation for the SET voltage (0.4 ± 0.03 V) is obtained, Figure S5(b). The $I_{ON}/I_{OFF}$ ratio for this device type was found to be $10^2$. The volatility of the memory has been studied from the retention time of the devices. The SET and RESET states were read with a sequential voltage pulse of 0.1V, with one second interval. The memory written in the Au top contact device was found to be volatile, with short retention times of ~500s, Figure 2(f). The addition of the Ag thin layer in between the Au and the VA-MoS$_2$, exhibit a memory retention of more than 20ks, Figure 2(g). The low ON state voltage (<0.5V) and high retention time in the Ag/VA-MoS$_2$/Si devices, demonstrate its robustness towards memristive applications.

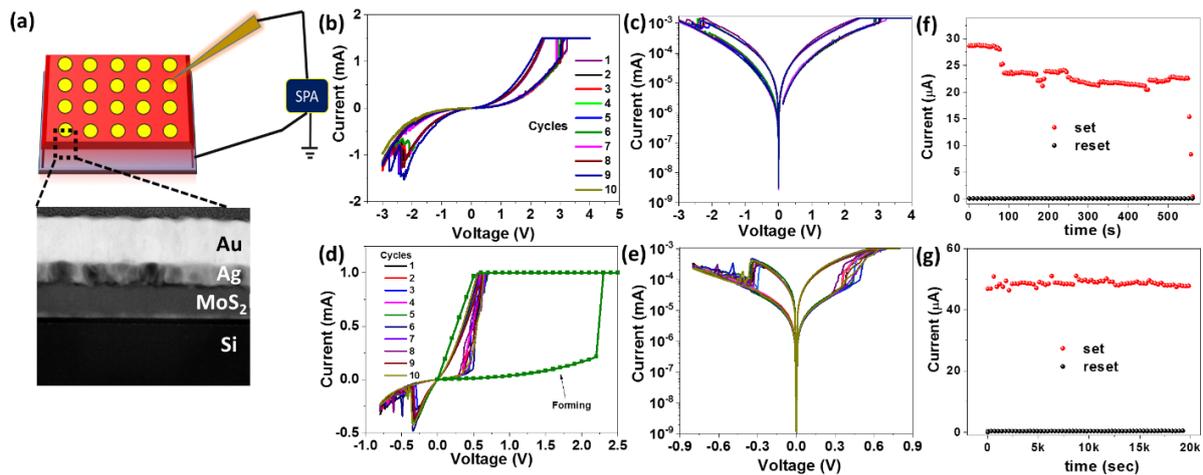

Figure 2: VA-MoS$_2$ based memristor device performance: (a) Schematic representation of the measurement set-up. (b) - (c) I-V sweeps for an Au/VA-MoS$_2$/Si memristor. (d) - (e) The same for a Au/Ag/VA-MoS$_2$/Si device. (f) and (g) Retention times as measured for the devices in (b) and (d), respectively.

High-angle annular dark-field (HAADF) STEM and elemental mapping were used to study and compare between the pristine and the turned "ON" devices. To this end, cross sectional samples were prepared via FIB methodologies (see experimental section) on both types of samples. STEM and EDS characterization of the pristine and ON-device are shown in Figures 3(a) and (b), respectively. No significant difference is observed in the HAADF STEM images, the top left in (a) and (b), for the pristine and set ON device, respectively. On the other hand, EDS of both samples show the presence of Ag in within the VA-MoS$_2$ film, only for the set-ON device (bottom right in (b)). This indicates that Ag ions/atoms diffused or intercalated between layers, though it may not necessarily be a continuous filament, forming the conducting path, as measured at the LRS.

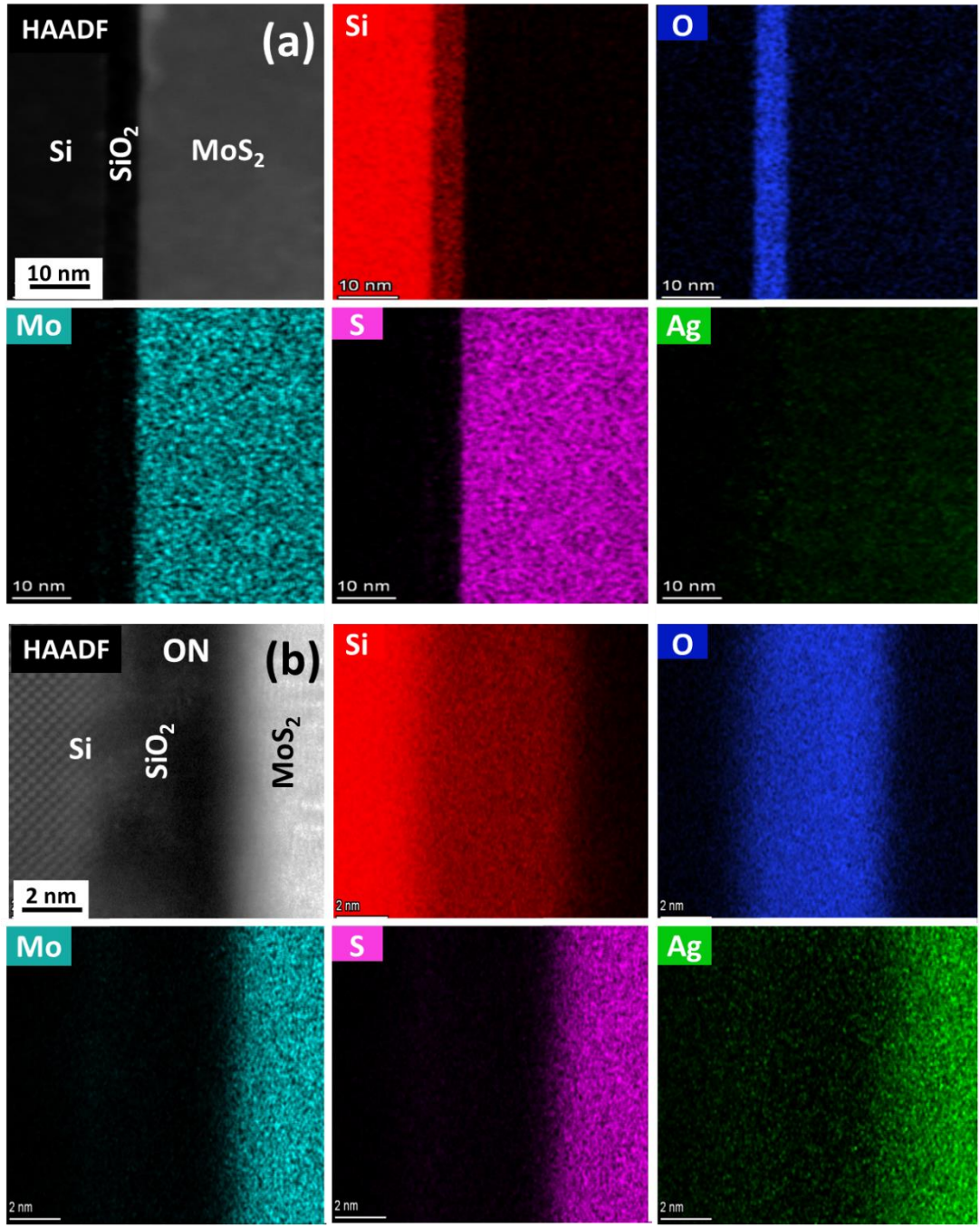

Figure 3: Cross sectional elemental EDS mapping of the memristive devices for pre-, (a), and post-, (b), forming. The set ON device, (b), shows the presence of Ag (green) in the VA-MoS$_2$ layer.

These observations are well correlated with the I-V profiles in the switching cycles. The I-V sweep of Ag/VA-MoS$_2$/Si devices exhibit bipolar characteristics with a close to linear characteristics at low voltages, Figures 2(c) and S6. Figure 4 (a) shows the ln (I) vs ln (V) profile for HRS and LRS states in a positive cycle. The HRS follows a space charge limited current (SCLC) model with an Ohmic region (slope ~ 1) at low voltages. Further voltage increase leads to a field driven I ~ $V^2$ (slope ~ 2: child law)[2a, 23] region for transition from HRS to LRS. Here, the positive bias drives the electron/ Ag ion between the electrodes through the $MoS_2$ film, forming a conducting filament that drives the HRS to LRS transition. The ohmic behaviour (slope ~ 1.2) in LRS further approves such filament formation through the silver diffusion, creating conduction path for the low resistive 'on' state.[23a] A schematic representation of the proposed Ag-enriched VA-$MoS_2$ film – assisted conduction path formation mechanism is shown in Figure 4(b). In this scheme, the diffusion of the Ag ions/atoms (with ionic/atomic radii of ~0.129-0.160 nm) through the channels created by the vertically aligned $MoS_2$ layers, with a vdW gap of ~0.314 nm.[23b] First principle calculations also demonstrate that the electrochemical intercalation of Ag in vdW gap of the 2H-$MoS_2$ is energetically favourable with low activation energies.[24] The application of a positive bias to the Ag electrode leads to its oxidation into $Ag^+$ ions, facilitating their diffusion towards the bottom contact, the $Si^{++}$ substrate, for the reduction to take place. This phenomena is the reason for the conducting Ag-based filament-like in between the VA-$MoS_2$ layers, turning the device into the low resistance state. When the device polarity is reversed, the continuity of the Ag filament is disrupted due to reverse migration of the intercalated Ag atoms. Hence, the conduction properties seen in the sweeping cycles arise from the complex hybrid structure of Ag intercalated $MoS_2$.

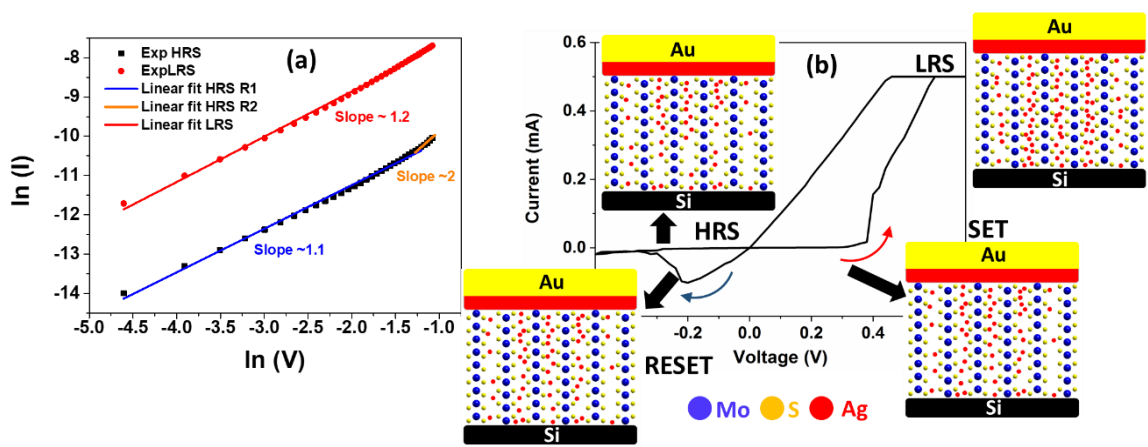

Figure 4: (a) ln (I) vs ln (V) show the SCLC characteristics in HRS with Ohmic and I α $V^2$ (child's law) regions. The LRS shown the slope of 1.2 while fitted with linear equation. (b) Schematic representation of the different states along the I-V cycle, while applying positive bias the HRS was transitioned to LRS with the diffusion of Ag ions. The continuity of the Ag ion disrupted and the device switched back to

HRS during resetting process. Note, in the reset state, part of the intercalated Ag ions/atoms stays throughout the VA-MoS$_2$ film.

The stability and robustness of the VA-MoS$_2$ memristor was tested as a function of temperature. In this scheme, the device is first turned ON to its LRS at room temperature (RT-SET) and then heated to 350 °C with a heating rate of 5° C/min. The device retained the LRS state, while current increases with temperature. The I-T curve of this RT-SET state was well fitted with a linear equation with a positive slope, Figure 5(a). In a complimentary experiment the device was switched ON at 350 °C (HT-SET), and then cooled down to RT, exhibiting a negative linear slope in the I-T curve, Figure S7(a), *i.e.* the current decreases while cooling. The direct relationship of the current with respect to temperature indicates a non-metallic behaviour in the LRS of the memristor.[33] This means that the diffusion/intercalation of Ag ions/atoms throughout the VA-MoS$_2$ film increases with the temperature, and accordingly, the current. Therefore, the LRS is achieved when the amount of Ag ions/atoms is enough to allow for the charges to hop efficiently through the silver-enriched VA-MoS$_2$. Figure 5(b) shows the Arrhenius plot, ln (I)-1/T, for the heating, Figure 5(a), and cooling, Figure S7(a), measurements. Interestingly, the activation energies, E$_a$, extracted from such plots in the heating set of measurements, (0.6 eV), was found to be larger than for the cooling process (0.35 eV). Such activation energies, imply the presence of shallow traps, suggesting the low resistance state is reached by trap-assisted charge transport, rather than the formation of a continuous metallic filament.[25] The difference in the E$_a$ can be understood according to the working mechanism principle mentioned above, in which the temperature enhances the diffusion of silver ions/atoms into the MoS$_2$ film, thus creating conductive centers that facilitate charge hopping, and hence the lower E$_a$ for the cooling set of measurements.

The memristor shows bipolar switching characteristics in all tested temperatures. It is found that the set voltages significantly decrease with the temperature raise. For 100 °C, 200 °C, 300 °C and 350 °C the set voltages are 0.5V, 0.35V, 0.27V and 0.2V, respectively, Figure 5(c). It reveals that along with applied potential between electrodes the temperature also assist to switch on the device. The comparison between pre- and post-heating testing cycles, Figure 5(d) show the permanent changes in the resistance of a device in both LRS and HRS states. It is also noted that cooling – heating cycle (figure S7(b)) does not yield changes in the initial resistance of the tested state of the memristor. Hence, low voltages (~10 mV) used for "reading" the state does not have an impact on the current. On the other hand, writing a memory (larger SET-V, ~0.4V) at particular temperature yield a permanent change in the resistance. In a memory cell the disruption occurs locally at the weakest part of the filament.[34] According to the temperature-dependent memristive results, the increased intercalation of Ag in between vdW gaps may not be entirely reversible.[26] Therefore, once the device set at a high temperature, the resistance of the Ag-MoS$_2$ (R$_1$-see SI for details) reduced permanently. Retention time is almost not affected by the operation temperature, as shown in Figure 5 (e), and the switching endurance of the device tested for more than 100 cycles (at room temperature, RT, and 350 °C), is presented in Figures 5 (f) and S8. Here, the endurance of the device was extracted from the current measured at 0.2 V in HRS and LRS of the continuous sweeping cycle.

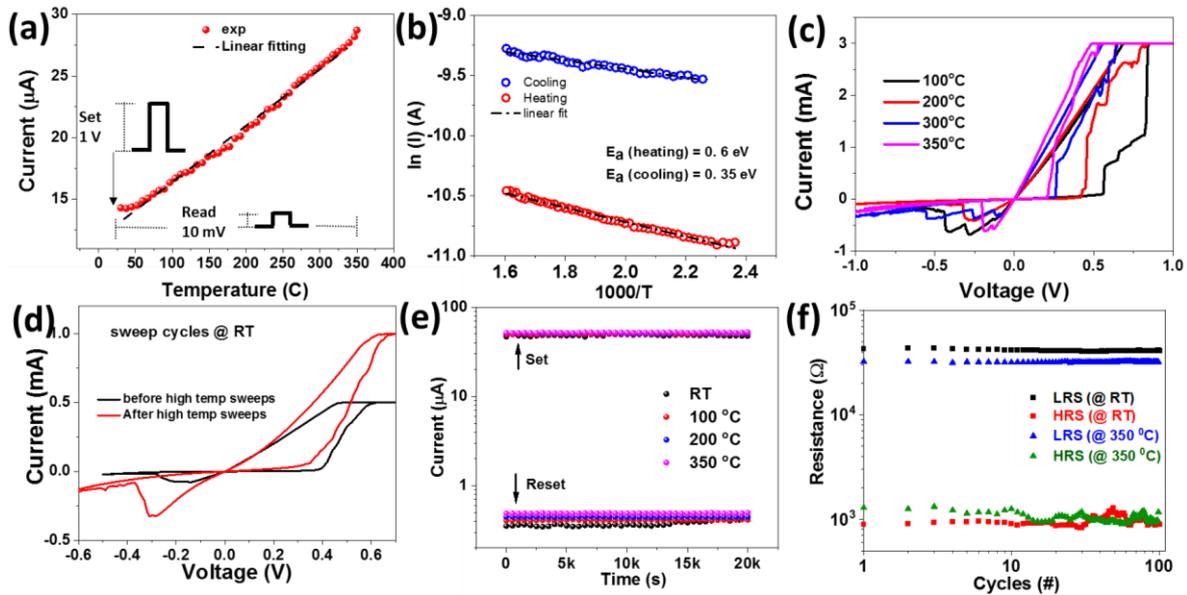

Figure 5: Thermal stability of the memristive devices at elevated temperature: (a) The device was set-ON at RT (RT-SET). The stored memory (set state) was tested up to 350 °C. The current increases with the temperature and fitted with linear equation. (b) Arrhenius plot, ln(I)-1/T, from the measurements in (a) and Figure S7(a), showing the extracted E$_a$=0.6 and 0.35 eV, for the heating and cooling measurements, respectively. (c) Switching characteristics at different temperatures. (d) RT and High temperature cycle operation showing the permanent reduction in the resistance of the device. (e) Retention time as a function of operation temperature. (f) Switching endurance at room temperature and at 350 °C.

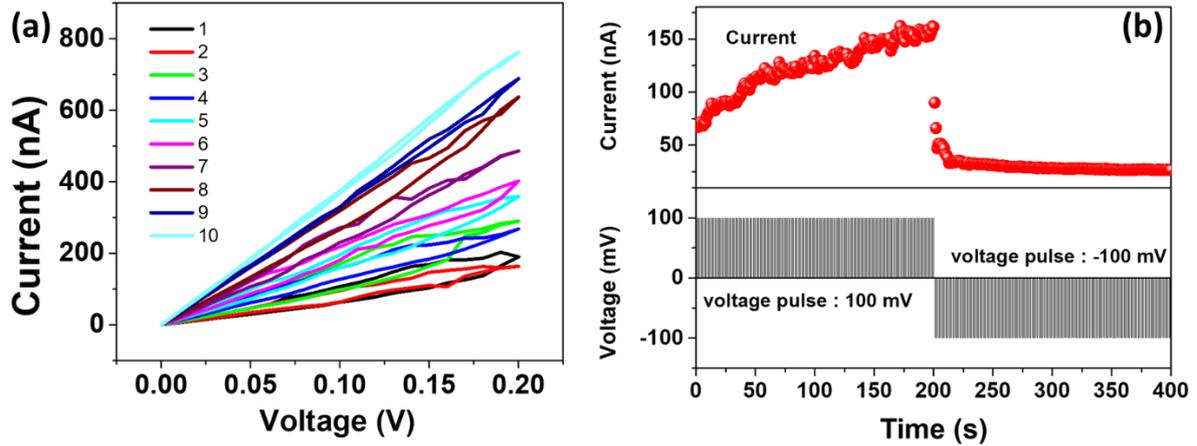

Figure 6: Weighted synaptic behavior: (a) Continuous positive I-V (V<V-set) sweep showing the monotonous conductance increment for each sweeping cycle. (b) Applied positive (100 mV) and negative (-100 mV) continuous pulses with respect to output current results in exponential current increase and decrease, manifesting long-term potentiation (LTP) and long-term depression (LTD) process, respectively.

The VA-MoS$_2$-based memristive devices were tested for emulating synaptic behaviour to be applied in neuromorphic computing. The I-V behaviour of ten continuous positive voltage sweeping cycles were tested, Figure 6(a), in the HRS regime (low voltages, < SET V). An hysteresis loop for each cycle is clearly seen, Figure 6(a). Moreover, the consecutive sweeping cycles increases the conductance of the device. This behaviour indicates the possibility of monotonous and sequential modulation of the intrinsic conductance of the memristive devices to simulate synaptic behaviour.[3b, 4, 27] We further investigate their weighted synaptic behaviour by applying voltage pulses. Here, the stress and depress of the neural synapsis were analysed by continuous positive and negative pulses, respectively. The voltage pulse train consists of 200 positive (100 mV) and 200 negative (-100 mV) pulses, Figure 6(b). The consecutive positive pulses stressing the device for long term potentiation (LTP) which are observed through the increases in current, Figure 6(c). Similarly Long term depression (LTD) are seen from exponentially reduced current through successive negative pulses. It is important to note that this modulation was observed for currents lower than 120 nA and a low power consumption of 7 nW (P = $V_{read}$ * I). This current is significantly lower than the ones reported in multilayer and polycrystalline VdW materials.[11c, 11e, 28] Further the presence of the voltage amplitude dependent potentiation also demonstrated by comparing 100 mV, 150mV and 200 mV voltage training pulses (Figure S9). These are required for multi-level memory devices which is used in simulating a weighted synaptic behaviour. The non-linear relation between the current and the pulse number is believed to arise from non-uniform conductance changes in response to identical pulses. Such changes could be related to the different conductive path or filament formation stages, and has been observed in metal-oxide based memristors.[29] This phenomena is unwanted because in order to keep a linear current modulation with the pulse number, as desired in neuromorphic computing,[29] the pulse amplitude/width must be changed accordingly, increasing the complexity of the whole circuit. The reason for the non-linearity behaviour in the above devices is not clear at this point and further work needed to be done on this topic. With a proper material (VA-TMDC) engineering, for example by creating vertical heterostructures or a gradient of alloying/doping elements (like a functional gradient material, FGM), the kinetics of the conducting path could be tuned and thus, the LTP and LTD behaviour.

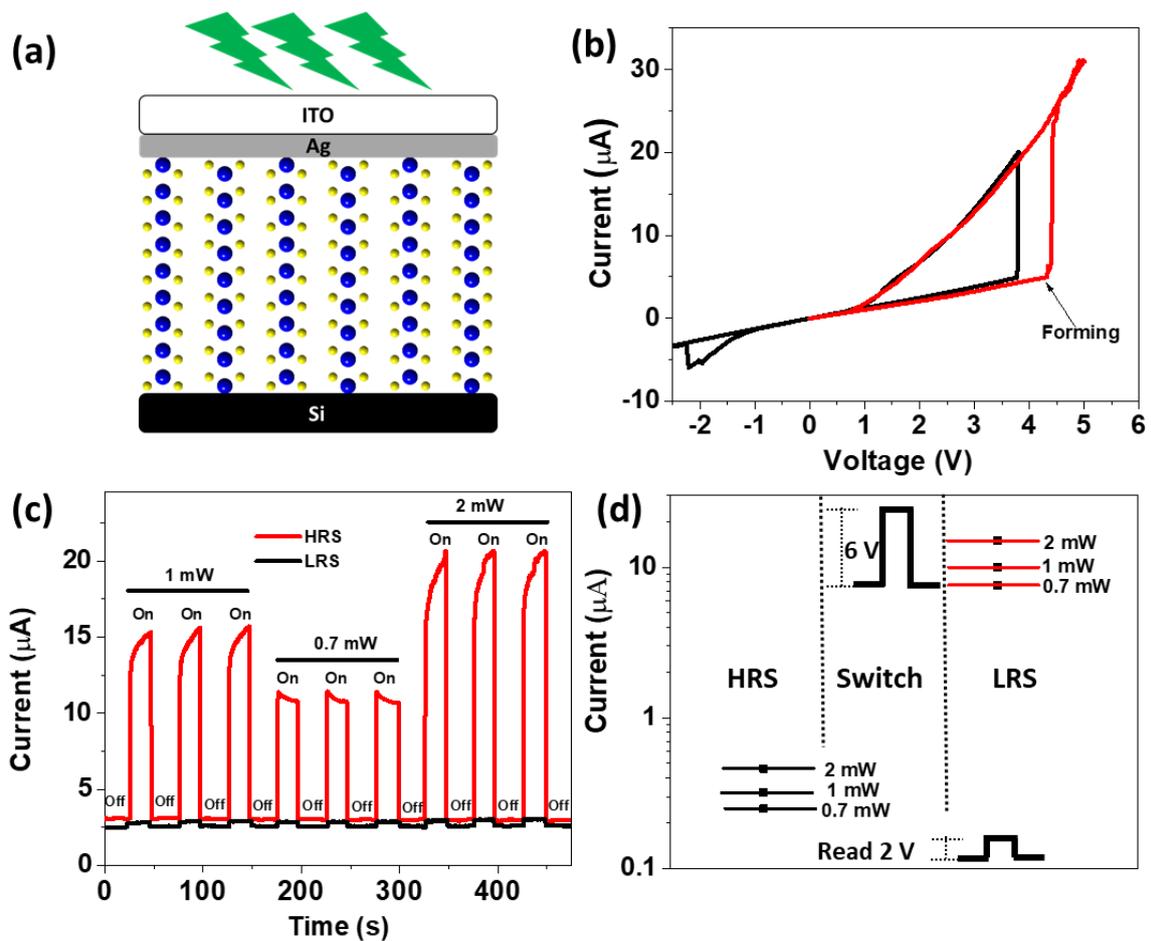

Figure 7: (a) Schematic representation of the transparent electrode based ITO/Ag/VA-MoS$_2$/Si device. (b) Electroforming and I-V sweep cycle. (c) Current enhancement in both states, HRS and LHS, via

light stimulation with different laser power intensities. (d) An example of six different memory states obtained in such device.

Along with the electronic properties, MoS$_2$ is also well known to be photosensitive,[8a] which has potential for integration with memristive applications,[16a] therefore, presenting an advantage over the more traditional metal-oxide based resistive switching materials. To study the photoactive characteristics of the VA-MoS$_2$ memrisors, the Au top contact was replaced with a transparent and conductive ITO thin layer. A device with the structure ITO (70 nm)/Ag (10 nm)/VA-MoS$_2$/Si was fabricated, as depicted in Figure 7(a). Here, ITO has been chosen as the transparent conducting electrode. A very thin Ag layer ensures the transmission between 50~60% of the incident visible light (here we used 532 nm),[30] to allow for testing the photoresponse of the memristor. The device exhibits an electroforming around 4.5V and the I-V sweeping cycle shows a significantly higher set voltage (3.5 V) than the Au/Ag /VA-MoS$_2$/Si devices (~0.4 V), Figure 7(b). The observed characteristics can be due to the thinner Ag layer used here in order to keep optical transparency. Further work is needed in order to optimize such device configuration. Nevertheless, the effect of light-induced current in both HRS and LRS can be clearly seen in Figure 7(c), where illumination intensity dependent current as a function of time characteristics of the device was studied. A constant potential of 2V was applied between the electrodes and a 532 nm laser used for illumination. The I-t curve shows that in both states, HRS and LRS, the current output increases upon illumination, Figure 7(c). The responsivity was calculated using the formula, $Responsivity = \frac{I_L - I_D}{P_{laser}}$. Here, I$_L$ and I$_D$ are the current measured under light and dark condition and P$_{laser}$, the incident laser power. The responsivity was found to be higher for the LRS comparing to HRS (Figure S(10)). In LRS the photoactivated charges might reach the bottom electrode due to the enhanced charge transport in the 'ON' state of the device. In both states, the photoresponse of the device was linear within the studied regime, Figure S10. Figure 7(d) shows an example in which six discrete states of the device were programmed, using three different laser power intensities, Figure 7(d). The photo-responsivity calculated for the LRS is found to be higher (>30 times) than for the HRS (Figure S10). The conduction path is formed by the diffusion of Ag ions/atoms into the VA-MoS$_2$, and thus, the low resistance state is achieved when higher amounts of silver is embedded in the film. The enhanced photoresponse in the LRS may indicate that the Ag contribute by enhancing the absorption[31] and /or facilitating the charge separation of the excited electron-hole pair upon illumination.[32] The light induced photo-memristive behavior has high potential to be implemented in high speed resistive switching devices,[15] as required for various applications.

**Conclusions**

In summary, a facile and robust approach is presented for the synthesis and assembly of Ag/VA-MoS$_2$/Si two-terminal memristor devices. The VA-MoS$_2$ film is grown by the sulfurization of pre-deposited Mo thin films and therefore, suitable for large-scale applications on various target substrates. The memristive device shows large retention times with low-SET voltages. Such devices exhibit reproducible and non-volatile behaviour at high operating temperatures, up-to 350°C, which, to our knowledge, is the highest reported so far (for 2D materials-based memristors). Weighted synaptic behaviour with long-term potentiation and depression (LTP and LTD, respectively) was demonstrated as well. Finally, the Ag/VA-MoS$_2$/Si devices exhibit photoresponse and hence, light-sensitive memristive behaviour is shown and, by changing the laser power used for the light stimulation, different memory states programmed. The facile and high scalability of the synthetic process used to make the VA-MoS$_2$, together with the high performance at RT and high temperatures, the synaptic behaviour and photo-sensitivity, may allow for the integration of large-scale multi-functional memristive devices. The same methodology shown here is suitable to be applied to other TMDCs and their heterostructures, Moreover, Doping and alloying strategies can be easily implemented in this approach,[9a, 14] therefore, opening a large window for property tunability.

**Acknowledgements**

The authors gratefully acknowledge the very generous support from the Israel Science Foundation, projects # 2171/17 (K.R.), 2549/17 (M.F.) and 1784/15 (A.I.).

**Experimental Section**

**Synthesis of VA-MoS2**

A heavily doped Si wafer (p$^{++}$) was treated with HF (30%) for 5s to etch native SiO$_2$ on Si. Immediately after etching the Si was loaded into e-beam evaporator (VINCI) for a coating of 15 nm Mo thin film. The Mo/Si were used for growing MoS$_2$ through sulfurization process in atmospheric chemical vapour deposition (CVD) system (Figure (S7)). The three step CVD process starts with annealing under 100 sccm of Ar and 5sccm H$_2$ at 200°C for 1hr. In next step, the gas flow was changed in to 60sccm Ar and 5sccm $H_2$. The furnace was ramped to 800°C with the rate of 6 °C/min and the sulphur powder was heated to 180°C by external heater. During this 2 hr of VA-$MoS_2$ growth time, the sulphur vapour (above its melting point, which is 120°C) was carried downstream into the reaction zone by the carrier gas. At the end of this process the furnace cools down naturally to room temperature and the sulfur heater is turned off when the furnace cools to the temperature of 600°C.

**Structural and chemical characterization**

The TEM lamellas were prepared by the lift-out method in a dual-beam focused ion beam (FIB; FEI Helios NanoLab Dual Beam G3 UC). Samples were thinned to 50 nm using a Ga+ ion beam at 30 kV with beam currents ranging from 0.43 nA down to 24 pA. Amorphization layer removal was carried at 5 kV and 2 kV, with beam currents of 15 pA and 9 pA, respectively. The high resolution transmission electron microscopy (HRTEM) with a probe corrected FEI/ThermoFisher Titan Cubed Themis G2 60-300 operated at 200keV acceleration voltage were used to acquire the microgram. The EDS maps were acquired using a Bruker Dual-X detector (Bruker Corp., Billerica, USA) and processed using the Thermo Fisher Velox software. The Raman spectra were collected with a green laser (532 nm) by HORIBA LabRAM HR Evolution Raman spectrometer. A 100× Olympus objective focuses the laser beam to a spot of <1μm diameter with 1800 gr $mm^{-1}$ grating. XPS measurements were performed in UHV (2.5x$10^{-10}$ Torr base pressure) using a 5600 Multi-Technique System (PHI, USA). The sample was irradiated with an Al Kα monochromated source (1486.6 eV) and the outcome electrons were analyzed by a Spherical Capacitor Analyzer using the slit aperture of 0.8 mm. Depth Profiling was done by means of $Ar^+$ Ion Gun sputtering (sputter rate was ~17 Å/min on /Si reference).

**Device fabrication and testing**

To fabricate metal-insulator-metal structure for a memristor the existing Si substrate was served as a bottom contact. The metal stacks Ag/ Au and Ag/ITO for top contact was coated by using e-beam evaporator (VST) and RF magnetron sputtering (PENTA sputter), respectively. While coating, a shadow mask was used to define the device area. The current-voltage (I‑V) characteristics of fabricated Si/$MoS_2$/Au structures were investigated in the dark and under laser excitation (532 nm) using a Keysight B1500A semiconductor device parameter analyzer. A Linkam probe stage (HFS350EV-PB4) was used to control temperature during I-V measurements.